\documentclass[twocolumn,showpacs,twoside,10pt,aps,superscriptaddress]{revtex4-2}
\usepackage{graphicx}
\usepackage{CJKutf8}
\usepackage{float}
\usepackage{subfigure}
\usepackage{epsfig}
\usepackage{epsf}
\usepackage{mathptmx}
\usepackage{amssymb}
\usepackage{amsmath}
\usepackage{amsthm}
\usepackage{multirow}
\usepackage{makecell}
\usepackage{float}
\usepackage{appendix}

\usepackage{hyperref}
\hypersetup{colorlinks=true,
            linkcolor=blue,
            anchorcolor=blue,
            citecolor=blue
            }

\begin{document}

\title{Direct Readout of Nitrogen-Vacancy Hybrid-Spin Quantum Register in Diamond by Photon Arrival Time Analysis}

\author{Jingyan He}
\affiliation{School of Physics \& School of Microelectronics, Hefei University of Technology, Hefei, Anhui 230009, China}
\affiliation{Research Center for Quantum Sensing, Zhejiang Lab, Hangzhou, 311000, China}

\author{Yu Tian}
\affiliation{Shenzhen Institute for Quantum Science and Engineering and Department of Physics, Southern University of Science and Technology, Shenzhen 518055, China}

\author{Zhiyi Hu}
\affiliation{School of Physics \& School of Microelectronics, Hefei University of Technology, Hefei, Anhui 230009, China}
\affiliation{Research Center for Quantum Sensing, Zhejiang Lab, Hangzhou, 311000, China}

\author{Runchuan Ye}
\affiliation{School of Physics \& School of Microelectronics, Hefei University of Technology, Hefei, Anhui 230009, China}
\affiliation{Research Center for Quantum Sensing, Zhejiang Lab, Hangzhou, 311000, China}

\author{Xiangyu Wang}
\affiliation{Shenzhen Institute for Quantum Science and Engineering and Department of Physics, Southern University of Science and Technology, Shenzhen 518055, China}

\author{Dawei Lu}
\email{ludw@sustech.edu.cn}
\affiliation{Shenzhen Institute for Quantum Science and Engineering and Department of Physics, Southern University of Science and Technology, Shenzhen 518055, China}

\author{Nanyang Xu}
\email{nyxu@zhejianglab.edu.cn}
\affiliation{Research Center for Quantum Sensing, Zhejiang Lab, Hangzhou, 311000, China}

\date{\today}% It is always \today, today,
             %  but any date may be explicitly specified
             
\begin{abstract}
Quantum state readout plays a pivotal role in quantum technologies, spanning applications in sensing, computation, and secure communication. In this work, we introduce a new approach for efficiently reading populations of hybrid-spin states in the nitrogen-vacancy center of diamond using a single laser pulse, which utilizes the excited state level anti-crossing mechanism at around 500 Gs. Reading spin state populations through this approach achieves the same outcome as traditional quantum state diagonal tomography but significantly reduces the experimental time by an order of magnitude while maintaining fidelity. Moreover, this approach may be extended to encompass full-state tomography, thereby obviating the requirement for a sequence of spin manipulations and mitigating errors induced by decoherence throughout the procedure.
\end{abstract}   
          
\maketitle

\section{Introduction}

Owing to its outstanding optical and electron-spin properties, the negatively charged nitrogen-vacancy (NV) center in diamond has emerged as a highly promising platform for solid-state qubits \cite{PhysRevX.6.041035,PhysRevX.4.031022,science.1155400,R.T.Harley_1984}. The NV center provides both an electron spin and a substitutional nitrogen nuclear spin ($^{14}\rm N$ or $^{15}\rm N$), and these spins are coupled via a hyperfine coupling \cite{science.aad8022,science.1231675,nature25781}. A notable feature of this system is its suitability for manipulation as a hybrid-spin quantum register. On one hand, the electron spin exhibits properties such as optical pumping and electron-spin-state-dependent fluorescence \cite{j.pnmrs.2016.12.001}. This enables feasible initialization and readout of the electron spin \cite{ncomms11526,mamin_nanoscale_2013}, even at room temperature \cite{oort_optically_1988}. On the other hand, nuclear spins offer significantly longer spin lifetimes in comparison to electrons \cite{TERBLANCHE2001107}, making them a valuable resource for applications such as quantum memories \cite{science.1139831,science.1220513,science.1131871}, computational nodes for quantum error correction \cite{cramer_repeated_2016}, and quantum communication \cite{reiserer_robust_2016,kalb_entanglement_2017}.

However, initialization and readout of nuclear spins with high fidelity remains a challenge due to the small magnetic moments of nuclear spins. To address this challenge, the electron spin often serves as an ancillary qubit, interacting with individual nuclear spins. Then the nuclear spin can be polarized by coherently transferring the state from the electron spin, and read out through the reverse process \cite{pfender_high-resolution_2019, lvovsky_iterative_2004, s41467-019-08544-z,hopper_spin_2018}. Achieving this necessitates the implementation of intricate quantum-control pulse sequences, operating at both microwave (MW) and radio frequencies (RF), to manipulate electron and nuclear spins, respectively. Nonetheless, due to its weak coupling to magnetic field, executing nuclear-spin operations takes roughly three orders of magnitude longer than the electron spin. As a result, controlling the nuclear-spin state is quite inefficient, and the microwave-induced heating effect hinders its application in thermally-sensitive sciences.

Due to the excited state level anti-crossing (ESLAC) phenomenon of NV center under a magnetic field around 500 Gs, the substitutional nitrogen spin can be automatically initialized via a dynamical nuclear polarization (DNP) process during the optical pumping \cite{PhysRevLett.101.117601,PhysRevLett.102.057403}. This mechanism is frequently employed as a standard NV hybrid-spin initialization protocol in various applications \cite{PhysRevLett.110.060502,pnas.1811994116}, while the readout method remains unchanged. In this work, we introduce a new avenue to directly obtain the spin state by analyzing the arrival time of the emitting photons in a single laser pulse. The difference in photon arrival time is also due to the ESLAC mechanism, which forms different photon time traces between the states. Combined with the ESLAC-based DNP protocol, we demonstrate a efficiency state initialization and readout experiment accompanied by exceeding an order of magnitude increase, this method can be extended to full-state tomography and reducing decoherence-induced errors.

\section{physical system}
\begin{figure}[t]
	\includegraphics[width=\columnwidth]{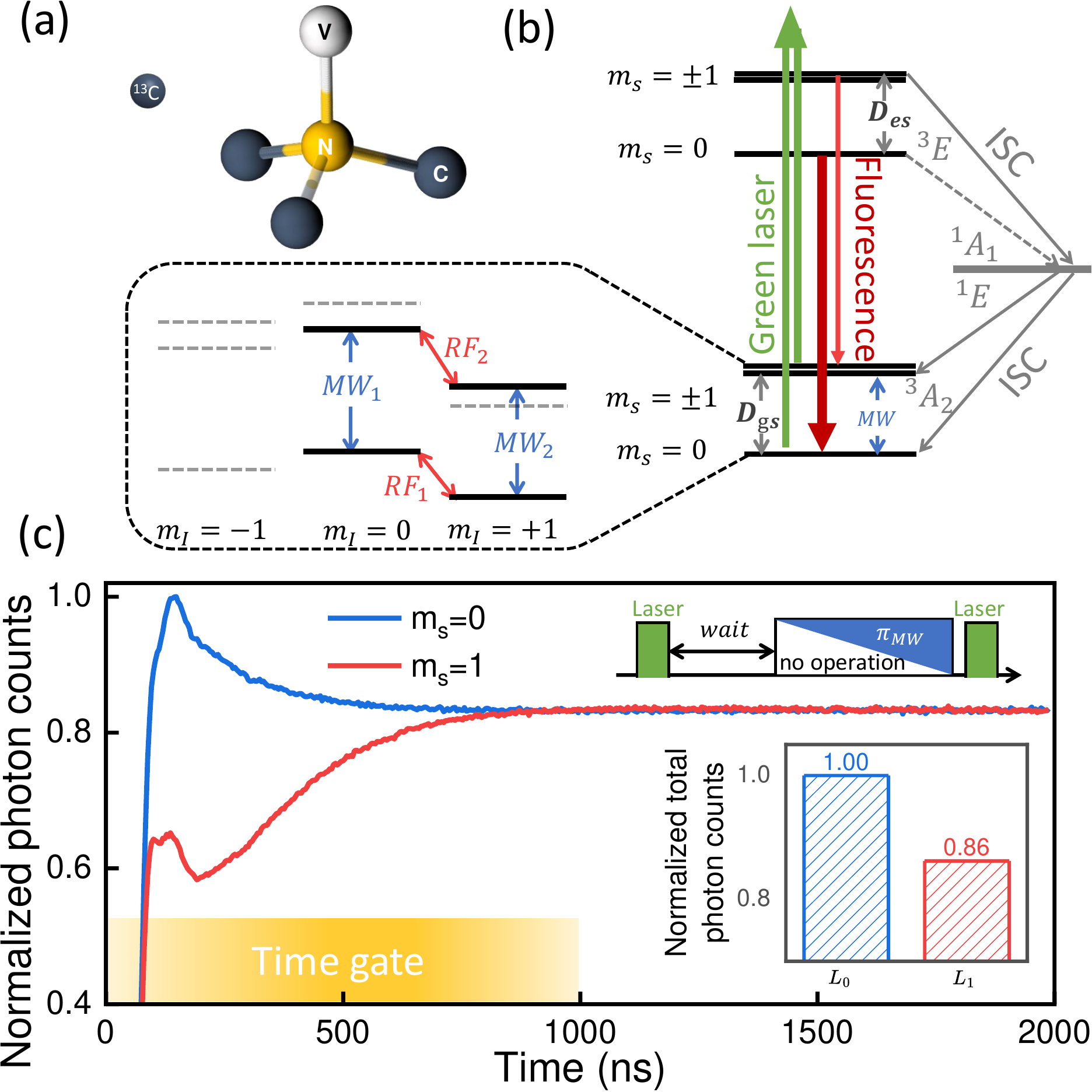}% Here is how to import EPS art
	\caption{\label{fig:fig1}(a) Atomic structure of the NV center in diamond. (b) Energy-level structure of the NV center. Optical transitions (green and red arrows), ISC (curved arrows) and MW transitions (blue arrows) are given. Dashed lines illustrate weak transitions. The intersystem crossing process ultimately results in the transfer of the spin state into $m_s=0$. The left panel illustrates the hyperfine interaction of the $\rm ^{14}N$ nucleus. The blue and red arrows correspond to the transitions of the electron spin $m_s=0\leftrightarrow m_s=-1$ and the nucleus spin $m_I=0\leftrightarrow m_I=+1$, respectively. (c) Photon time traces of the electron spin states, initialized at $m_s=1$ (in red) and $m_s=0$ (in blue). Inset: the normalized total photon counts of two initialized states $m_s=0$ $(L_0)$ and $m_s=1$ $(L_1)$. (d)  Calibration of fluorescence intensity of four pure two-qubit states. In this sequence, we employ ${\pi}_{MW_1}$, ${\pi}_{MW_2}$, ${\pi}_{RF_1}$, and ${\pi}_{RF_2}$, which represent $\pi$-rotations applied to the respective transitions. }
\end{figure}

The NV center in diamond is composed of a nitrogen atom substitution for a carbon atom, with a neighboring carbon vacancy as illustrated in Fig. \ref{fig:fig1}(a). There exist two charge states, namely the neutral ($\rm NV^0$) and the negatively charged ($\rm NV^-$) NV color centers \cite{PhysRevA.106.033506}. This investigation is primarily concerned with the negatively charged NV color center, characterized by a ground state in the form of a spin triplet state denoted as $^3A$. This state displays a zero-field splitting of $2.87$ GHz between its spin sublevels, specifically $m_s=0$ and $m_s=\pm1$. Additionally, there is an excited state denoted as $^3E$, also a spin triplet, with a zero-field splitting of $1.4$ GHz. The energy level diagram of the NV center is depicted in Fig. \ref{fig:fig1}(b).
The NV center in diamond demonstrates a spin-dependent cycling transition of the electron spin under the influence of laser excitation \cite{pssa.200671403,j.pnmrs.2016.12.001,science.1131871}.

In Fig. \ref{fig:fig1}(b), the spin state $m_s=0$ exhibits a notably low probability of undergoing intersystem crossing (ISC) \cite{PhysRevLett.114.145502,j.pnmrs.2016.12.001} and a heightened likelihood of experiencing excitation to the corresponding $m_s=0$ state within the $^3E$ state (with a lifetime of $12$ ns) before reverting to its initial state. Conversely, for the spin state $m_s=\pm1$, there is a substantial likelihood of transitioning to the intermediate $^1A$ state (with a lifetime of $250$ ns), and an excited state lifetime of $7.8$ ns for $m_s=\pm 1$. Once the NV center is in its metastable state, the decay back into the triplet ground state preferably occurs into the $m_s=0$ state with high fidelity. Under optical pumping, the electron spin state will ultimately reach $m_s=0$, regardless of its initial state.

Although laser illumination initializes the electron spin state, studying fluorescence evolution over time for each state (see Fig. \ref{fig:fig1}(c)) is insightful. If the electron spin state was $m_s=0$ before laser illumination, it remains there, continually undergoing optical excitation and fluorescence emission cycles. This results in a sustained high level of fluorescence, except initially, where a small ISC rate, non-luminous and long-lived, leads to the initial fluorescence decrease. However, if the spin state was $m_s=\pm 1$, the higher ISC rate causes the electron spin to eventually transition from the excited state triplet to the excited singlet state during laser illumination. This transition leads to a breakdown in the fluorescence count rate until the NV returns to its ground state with $m_s=0$.

 The crucial aspect for optical readout of the electron spin state is that the NV center in the $m_s=0$ state emits a higher number of photons within the first few hundred nanoseconds compared to when it is in the $m_s=\pm1$ state. This leads to approximately $30 \%$ more total photon counts, which we refer to as the ``signal photon''. This difference in emitted photons for $m_s=0$ and $m_s=\pm1$ states allows for the electron spin state of the NV center to be efficiently read out at room temperature \cite{s11467-022-1235-5,suter_single-spin_2017,gulka_room-temperature_2021,warren_usefulness_1997,song_pulse-width-induced_2020}. The two polarized electron spin states, $m_s=0$ and $m_s=1$, correspond to the upper and lower fluorescence boundaries of the NV system, denoted as $L_0$ and $L_1$, respectively. Any superposition of these two states results in a general photon count $L$ that falls between these two fluorescence boundaries when measured repeatedly. $L$ can be expressed as a function of $L_0$ and $L_1$ by $L=c_0 L_0+c_1 L_1$, 
where $c_0$ and $c_1$ denote the respective probabilities of the states $m_s=0$ and $m_s=1$, and $c_0+c_1=1$. Consequently, the electron spin state can be ascertained through the measurement of fluorescence counts.

The achievement of single-shot readout for an individual nuclear spin has been successfully demonstrated utilizing a single NV center in diamond \cite{science.1189075,PhysRevLett.118.150504,gillard_harnessing_2022,PhysRevApplied.12.024055}. This work was realized within the NV hybrid-spin system, where an NV center is coupled with $\rm ^{14}N$ nuclear spins, forming a spin-triplet state. The hybrid-spin system is illustrated in Fig. \ref{fig:fig1}(b). When a magnetic field $B$ is applied along the NV defect axis, the Hamiltonians for both the ground state and the excited state share the same form, given by $H=D\hat{S}^2_z+{\gamma}_eB\hat{S}_z+{\gamma}_nB\hat{S}_z+A\hat{S}\hat{I}$. Here, $D$ represents the zero-field splitting, ${\gamma}_e$ and ${\gamma}_n$ denote the gyromagnetic ratios for the electron spin and nuclear spin, respectively, while $\hat{S}$ and $\hat{I}$ correspond to the operators for the electron and nuclear spins. Additionally, $A$ signifies the strength of the hyperfine interaction. At a magnetic field strength of 500 Gs, corresponding to the ESLAC region, electron-nuclear-spin flip-flops occur, leading to polarization into the $m_s=0$ state with $m_I=+1$.

Although the electron spin is a spin-1 system, a subspace containing only $m_s = 0$ and $m_s=+1$ (or -1) is utilized to form a two-level quantum bit (i.e., a qubit) in most cases of quantum information processing \cite{PhysRevLett.123.183602,PhysRevX.9.031045}. Specifically for quantum sensing, a spin-1 state ($m_s=\pm1$) has a relatively higher mangnetic moment, which can enhance magnetic sensing performance. However, the readout of $m_s=+1$ and $m_s=-1$ states still relies on the help of $m_s=0$ state \cite{PhysRevLett.113.030803,PhysRevLett.118.197201,PhysRevX.8.031025}. Here, we consider only the subspace of $m_s=0$ and $m_s=-1$ within the ground spin-triplet state and denote the states with $m_s=0$ and $m_s=-1$ within the ground spin-triplet state as $\left|0\right\rangle_e$ and $\left|1\right\rangle_e$, respectively, corresponding to the electron spin qubit. For the nuclear spin qubit of the $\rm ^{14}N$ nuclei, we use $\left|\uparrow\right\rangle_n$ and $\left|\downarrow\right\rangle_n$ to represent the states with $m_I=0$ and $m_I=+1$, as depicted in Fig. \ref{fig:fig1}(b). The resulting polarized spin state achieved in this system is $\left|0,\downarrow\right\rangle$.
In a similar vein, the total fluorescence counts, denoted as $L$, are directly related to the population of the four basis states i.e., $\left|0,\uparrow\right\rangle$, $\left|0,\downarrow\right\rangle$, $\left|1,\uparrow\right\rangle$, and $\left|1,\downarrow\right\rangle$ as follows:
\begin{equation}\label{equ2}
	L=c_{0\uparrow} L_{0\uparrow}+c_{0\downarrow} L_{0\downarrow}+c_{1\uparrow} L_{1\uparrow}+c_{1\downarrow} L_{1\downarrow},
\end{equation}
where $c_{0\uparrow}$, $c_{0\downarrow}$, $c_{1\uparrow}$ and $c_{1\downarrow}$ denote the respective probabilities of the corresponding states , and $c_{0\uparrow}+c_{0\downarrow}+c_{1\uparrow}+c_{1\downarrow}=1$.

However, Eq. (\ref{equ2}) is inadequate for the complete readout of the hybrid-spin states. To achieve spin state readout, the conventional approach involves the application of a set of unitary operations designed to transform the population, followed by another round of fluorescence count measurements. The specific experimental sequence is shown in Table \ref{table1}.

\begin{table}[ht]
\begin{tabular}{ccc} 
	\hline
	Readout Pulse&Transition&Counts\\
	\hline
	\hline
	no operations&no transform&$L_0$\\
	${\pi}_{MW_2}$&$\left|0,\downarrow\right\rangle \leftrightarrow\left|1,\downarrow\right\rangle$&$L_1$\\
	${\pi}_{RF_1}$&$\left|0,\downarrow\right\rangle \leftrightarrow\left|0,\uparrow\right\rangle$&$L_2$\\
	${\pi}_{MW_2}-{\pi}_{RF_2}-{\pi}_{MW_2}$&$\left|0,\downarrow\right\rangle \leftrightarrow\left|1,\uparrow\right\rangle$&$L_3$\\
	\hline
\end{tabular}
\caption{\label{table1} 
Experimental readout pulse sequence is designed to measure the various spin states within the electron-nuclear spin hybrid system. In this sequence, we employ ${\pi}_{MW_1}$, ${\pi}_{MW_2}$, ${\pi}_{RF_1}$, and ${\pi}_{RF_2}$, which represent $\pi$-rotations applied to the respective transitions, as illustrated in Fig. \ref{fig:fig1}. }
\end{table}

The experimental process described above can be succinctly summarized by the following matrix equation  
\begin{equation}\label{equ4}
	\left[\begin{array}{cccc}
		L_{0\uparrow}&L_{0\downarrow}&L_{1\uparrow}&L_{1\downarrow}\\
		L_{0\uparrow}&L_{1\downarrow}&L_{1\uparrow}&L_{0\downarrow}\\
		L_{0\downarrow}&L_{0\uparrow}&L_{1\uparrow}&L_{1\downarrow}\\
		L_{0\uparrow}&L_{1\uparrow}&L_{0\downarrow}&L_{1\downarrow}
	\end{array}
	\right]\cdot\left[
	\begin{array}{c}
		c_{0\uparrow}\\
		c_{0\downarrow}\\
		c_{1\uparrow}\\
		c_{1\downarrow}
	\end{array}
	\right]=\left[
	\begin{array}{c}
		L_0\\
		L_1\\
		L_2\\
		L_3
	\end{array}
	\right].
\end{equation}
Utilizing Eq. (\ref{equ4}), we can calculate the probabilities of different states and consequently determine the state being read out.

\section{Direct spin-readout scheme}

Here, our primary focus centers on the photon time trace, which offers a detailed account of how fluorescence evolves over time for each state. We have effectively harnessed this approach to successfully read out the electron spin state, as previously demonstrated \cite{qian_machine-learning-assisted_2021}.

\begin{figure}[H]
	\includegraphics[width=\columnwidth]{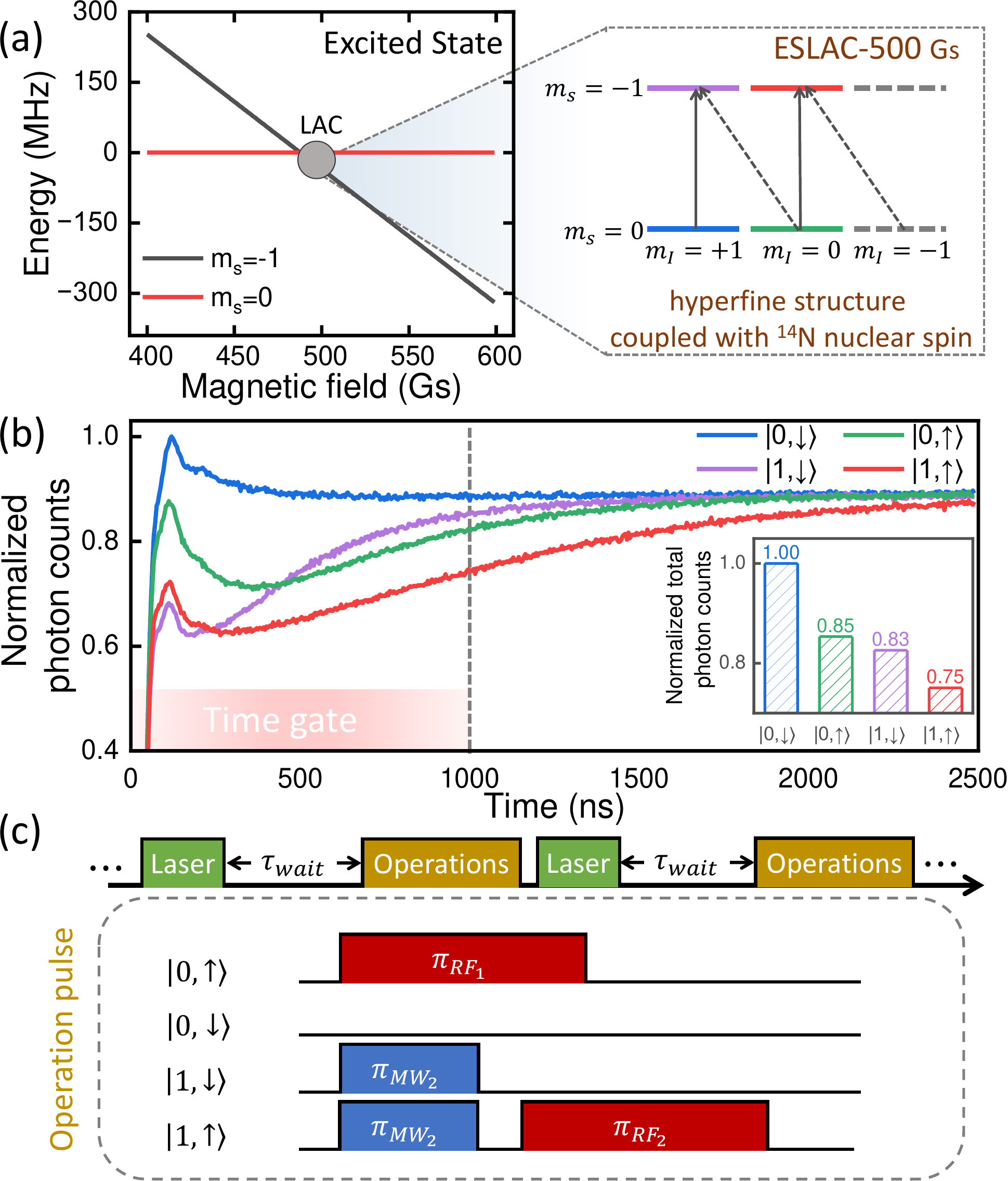}% Here is how to import EPS art
	\caption{\label{fig:fig2}(a) Eigenstates of the excited-state Hamiltonian displaying the ESLAC at $B\approx500$ Gs. Inset: the hyperfine structure coupled with the $\rm ^{14}N$ nuclear spin in the excited state. The gray solid and dashed arrows represent allowed and prohibited transitions, respectively. 
		(b) Photon time traces of spin states initialized at $\left|0,\uparrow\right\rangle$ (in green), $\left|0,\downarrow\right\rangle$ (in blue), $\left|1,\uparrow\right\rangle$ (in red), and $\left|1,\downarrow\right\rangle$ (in purple) at 500 Gs, respectively. Inset: the total number of photons within the time gate. (c) Inset(above): Readout sequence of electron spin; the laser readout process of the last experiment is the polarization process of the next experiment.  
	}
\end{figure}

At a magnetic field strength of 500 Gs, a noteworthy phenomenon called ESLAC emerges, as depicted in Fig. \ref{fig:fig2} (a). Since the state $\left|0,\uparrow\right\rangle$ and $\left|1,\downarrow\right\rangle$ exchanges in the excited states due to the ESLAC mechanism, the nuclear states become unbalanced and also been polarized in the same process \cite{PhysRevLett.102.057403,PhysRevA.80.050302,PhysRevB.81.035205,PhysRevB.85.134107}. Crucially, the transition from $\left|0,\downarrow\right\rangle$ to the excited state conserves the nuclear spin, resulting in sustained high fluorescence until a small ISC rate leads to fluorescence decrease. Meanwhile, the transition from $\left|0,\uparrow\right\rangle$ has a high probability of transitioning to  $\left|1,\downarrow\right\rangle$ with a high ISC rate, thus fluorescence rapidly drops until the spin in metastable state returns to the ground state of $\left|0,\downarrow\right\rangle$. Consequently, the hybrid-spin system in the state $\left|0,\downarrow\right\rangle$ emits a higher number of photons compared to the  state $\left|0,\uparrow\right\rangle$. By incorporating the optically polarized electron spin mechanism, as described in the second part, where the electron spin $m_s= \pm 1$ is lower than $m_s= 0$, it becomes evident that there are differences in fluorescence intensities among the four energy levels, namely $\left|0,\uparrow\right\rangle$, $\left|0,\downarrow\right\rangle$, $\left|1,\uparrow\right\rangle$, and $\left|1,\downarrow\right\rangle$.

The nuclear spin-dependent photon time traces are depicted in Fig. \ref{fig:fig2}(b), and these traces are acquired after activating the readout laser. They are recorded using a custom-built time tagger with a resolution of 2 ns, implemented on a commercial field-programmable gate array module \cite{ye_synchronized_2022}. If we assume there are $n$ time bins within the detection window, and $m^i$ (where $i$ ranges from 1 to $n$) represents the photon counts residing in the $i$-th time bin, then an arbitrary photon time trace can be expressed as a one-dimensional vector $M=\left[m^1, \cdots, m^i, \cdots, m^n\right]^T$. The fluorescence count $m^i$ obtained from the readout is intricately linked to the population of basis states as 
\begin{equation}\label{equ6}
    m^i=l^i_{0\uparrow}c_{0\uparrow}+l^i_{0\downarrow}c_{0\downarrow}+l^i_{1\uparrow}c_{1\uparrow}+l^i_{1\downarrow}c_{1\downarrow}.
\end{equation}
In this context, $l^i_{0\uparrow}, l^i_{0\downarrow}, l^i_{1\uparrow}$ and $l^i_{1\downarrow}$ denote the photon counts present in the $i$-th time bin for the four basis states, respectively. Since there are $n$ time bins, we can derive $n$ functions, which can be represented in the matrix form
\begin{equation}\label{equ7}
	\left[\begin{array}{cccc}
		l^1_{0\uparrow}&l^1_{0\downarrow}&l^1_{1\uparrow}&l^1_{1\downarrow}\\
		l^2_{0\uparrow}&l^2_{0\downarrow}&l^2_{1\uparrow}&l^2_{1\downarrow}\\
		\vdots&\vdots& \vdots&\vdots\\
		l^i_{0\uparrow}&l^i_{0\downarrow}&l^i_{1\uparrow}&l^i_{1\downarrow}\\
		\vdots&\vdots& \vdots&\vdots\\
		l^n_{0\uparrow}&l^n_{0\downarrow}&l^n_{1\uparrow}&l^n_{1\downarrow}\\
	\end{array}
	\right]\cdot\left[
	\begin{array}{c}
		c_{0\uparrow}\\
		c_{0\downarrow}\\
		c_{1\uparrow}\\
		c_{1\downarrow}
	\end{array}
	\right]=\left[
	\begin{array}{c}
		m^1\\
		m^2\\
		\vdots\\
		m^i\\
		\vdots\\
		m^n
	\end{array}
	\right].
\end{equation}

This function describes how an unknown state, represented as a vector $M=\left[m^1, \cdots\, m^i, \cdots\, m^n\right]^T$, can be expressed using a set of basis vectors $L_{0\uparrow,0\downarrow,1\uparrow,1\downarrow}=\left[l^1_{0\uparrow,0\downarrow,1\uparrow,1\downarrow},  \cdots, l^i_{0\uparrow,0\downarrow,1\uparrow,1\downarrow}, \cdots, l^n_{0\uparrow,0\downarrow,1\uparrow,1\downarrow}\right]^T$ within a Hilbert space. We use matrix form to represented the basis vectors as $L=\left[L_{0\uparrow}, L_{0\downarrow}, L_{1\uparrow}, L_{1\downarrow}\right]$. The coefficients $c=\left[c_{0\uparrow}, c_{0\downarrow}, c_{1\uparrow}, c_{1\downarrow}\right]^T$ correspond to the probabilities associated with the four coefficients.

As an alternative to linear inversion methods, maximum likelihood estimation has been widely employed \cite{lvovsky_iterative_2004,PhysRevA.61.010304}. In this work, we utilize optimization techniques to perform a direct analysis of the photon time trace, with the objective of determining the optimal values $c^{\rm optimal}$ that provide the best approximation to $c^{\rm theoretical}$ with respect to a specific matrix norm. The optimization process is subject to a constraint condition, which is defined as
\begin{center}
\begin{equation}\label{equ8}
        \begin{split}
		&min\qquad \|L\cdot c-M\|_2,\\
		&s.t.\qquad c^T \cdot c=1.
        \end{split}
\end{equation}
\end{center}

\section{Experimental Realization}
In our experimental setup, we conducted tests using an NV center embedded in a bulk chemical vapor deposition diamond. These experiments were carried out on a custom-built confocal microscopy system, all of which took place at room temperature. The specific NV center utilized in these experiments was positioned at a depth of 10 $\rm \mu m$ within the diamond material. This NV center was addressed and manipulated using our in-house optically detected magnetic resonance system. The experimental setup involves a 532 nm green laser beam that is passed through an acoustic-optic modulator to facilitate switching. After passing through an oil objective lens and focusing on the NV center, the fluorescence emitted is collected by an avalanche photo diode (APD). The output signal from the APD is then detected by a custom-built time tagger with a resolution of 2 ns. To create a static magnetic field, approximately 500 Gs, a columnar neodymium magnet is employed. This magnetic field is applied parallel to the NV axis, which is in the $z$-direction. The purpose of this magnetic field is to exclusively affect the $z$-terms of the electron spin $\rm \hat{S}_z$, leading to the splitting of the $m_s=\pm1$ sublevels and enabling the ESLAC. This ESLAC phenomenon facilitates the optical polarization of the nuclear spin with ease. To manipulate the electron spin and nuclear spin for general-purpose measurements, MW and RF sources are utilized, respectively. The frequencies of RF$_1$ (5.102067 MHz) and RF$_2$ (2.941124 MHz) have been calibrated through electron-nuclear double resonance experiments.

To validate the procedure, our initial step involved calibrating the photon time traces of the four basis states through a substantial number of measurements, totaling $10^9$. These calibrated traces are collectively represented as a set of basis vectors, denoted as $\left\{L_{0\uparrow}, L_{0\downarrow}, L_{1\uparrow}, L_{1\downarrow}\right\}$. Subsequently, we prepared test states, namely $\left|0,\uparrow \right\rangle, \left|0,\downarrow\right\rangle, \left|1,\uparrow\right\rangle$, and $\left|1,\downarrow\right\rangle$, through different operations. And the fidelity of test sate are over 0.99 (see Appendix \ref{app2}). We then carried out laser-based readouts for these test states, totaling $10^7$ measurements, and the results were documented as $M$.
Utilizing the optimal method, we successfully obtained the four coefficients: $c_{0 \uparrow}, c_{0 \downarrow}, c_{1 \uparrow}, c_{1 \downarrow},$ and readout the spin state populations with fidelity 0.9972 $\pm$ 0.00042, 0.9963 $\pm$ 0.00043, 0.9981$\pm$ 0.00038, and 0.9982 $\pm$ 0.00038, respectively. Here, we characterize the fidelity of population readout as Equ. (\ref{equ20}), obtained from the full state tomography caculation \cite{PhysRevLett.130.090801,j.physleta.2008.10.083}.
\begin{equation}\label{equ20}
	F_p=\frac{(c^{\rm th}, c^{\rm exp})}{\sqrt{(c^{\rm th}, c^{\rm th})(c^{\rm exp}, c^{\rm exp})}},
\end{equation}
where $c^{\rm th}$ ($c^{\rm exp}$) is theoretical( experimentally reconstructed) population vectors formed by $\left[c_{0\uparrow}, c_{0\downarrow}, c_{1\uparrow}, c_{1\downarrow}\right]^T$.

Now, we conside the population readout of an arbitrary state, in Fig. \ref{fig:fig2}.(b) the time trace of basis state are calibrated and we think the time trace of arbitrary state is straightforwardly the superposition of the four basic states, which has be verified in Supplementary Materials. Here, we perform one more experiment employing the superposition states of electron or nuclear spins and the results are shown in Table \ref{table2}. 
 
\begin{table}[h]
\resizebox{1\columnwidth}{!}{	
\begin{tabular}{cccccccccccc}
	\hline
	&$\frac{1}{\sqrt{2}}(\left|0,\uparrow\right\rangle+\left|0,\downarrow\right\rangle)$&$\frac{1}{\sqrt{2}}(\left|0,\downarrow\right\rangle+\left|1,\downarrow\right\rangle)$&$\frac{1}{\sqrt{2}}(\left|1,\uparrow\right\rangle+\left|1,\downarrow\right\rangle)$\\
	\hline
	\hline
	$c_{0\uparrow}$&0.42180$\pm$0.00320(0.5)&0.04460$\pm$0.00087(0)&0.09760$\pm$0.00011\\
    $c_{0\downarrow}$&0.51137$\pm$0.01311(0.5)&0.52938$\pm$0.01486(0.5)&0.00000$\pm$0.00914\\
    $c_{1\uparrow}$&0.05892$\pm$0.00047(0)&0.00000$\pm$0.00058(0)&0.44625$\pm$0.00472\\
	$c_{1\downarrow}$&0.00791$\pm$0.00310(0)&0.42602$\pm$0.00807(0.5)&0.45616$\pm$0.00949\\
	\hline
\end{tabular}}
\caption{\label{table2} 
Measured coefficients of superposition states using the photon time trace method. }
\end{table}

The obtained fidelity values, denoting the similarity between the input states $\frac{1}{\sqrt{2}}(\left|0,\uparrow\right\rangle+\left|0,\downarrow\right\rangle), \frac{1}{\sqrt{2}}(\left|0,\downarrow\right\rangle+\left|1,\downarrow\right\rangle)$ and $\frac{1}{\sqrt{2}}(\left|1,\uparrow\right\rangle+\left|1,\downarrow\right\rangle)$ are  0.99145 $\pm$ 0.00065, 0.99206 $\pm$ 0.00050 and 0.98845 $\pm$ 0.00083 respectively. The observed differences between the experimental and theoretical results can primarily be attributed to factors such as statistical variations in photon detection and control errors arising from pulse imperfections during state preparation and tomography implementation.

\section{time-cost analysis}
The direct readout method sets itself apart from the conventional approach by eliminating the need for a spin operation sequence. In the general method, reading out the nuclear spin state of the NV hybrid-spin quantum register necessitates at least four operations and laser-based readout. However, given the effective isolation of nuclear spins from their surrounding environment, manipulating nuclear spins entails executing a lengthy RF sequence lasting approximately 100 $\rm \mu$s.

\begin{figure}[H] 
	\includegraphics[width=0.5\textwidth]{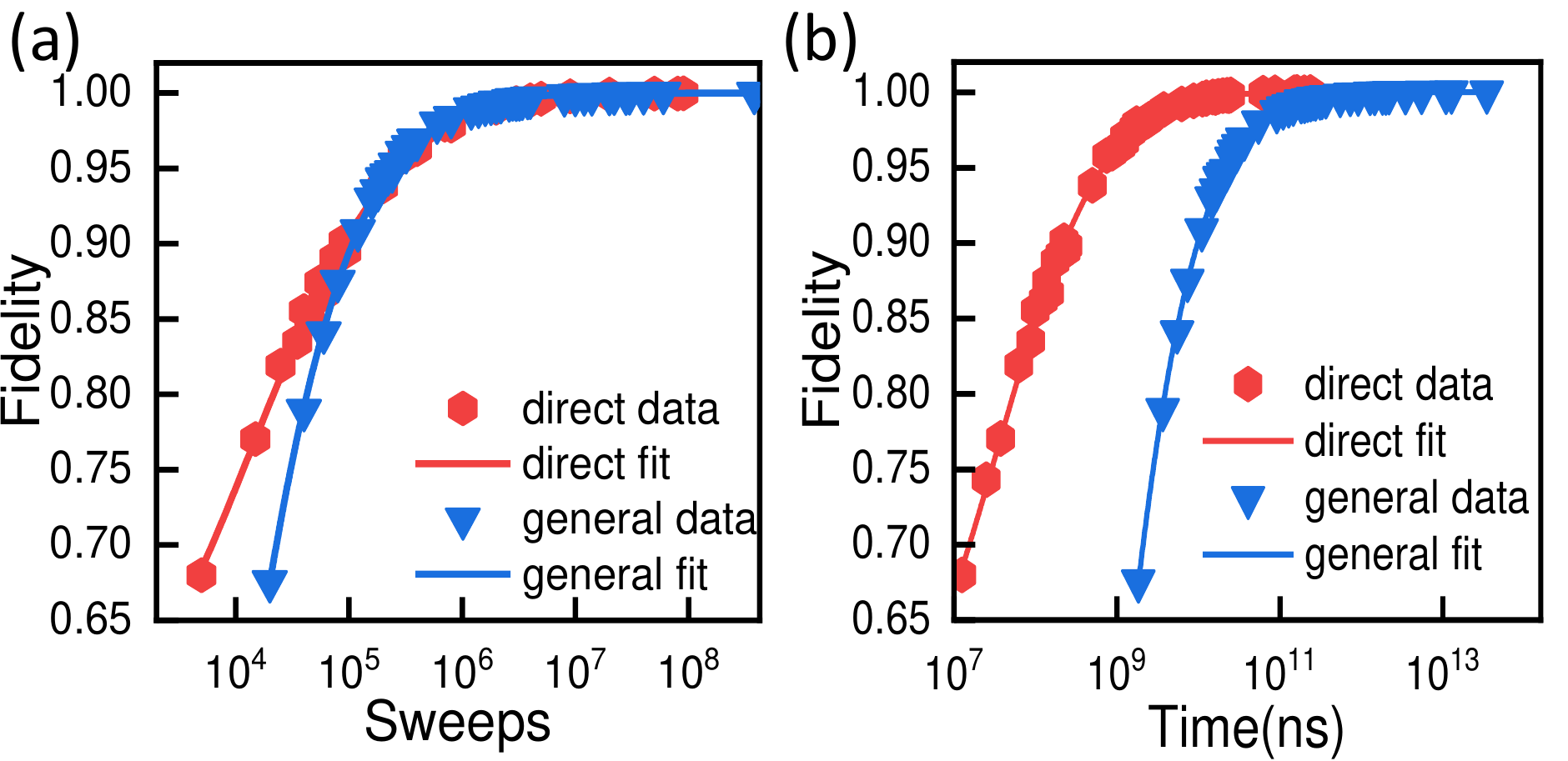}
	\caption{\label{fig:fig3} (a) Evolution of fidelity $F_p$ as a function of sweeps under 500 Gs. The data for both methods are well-fitted using the function $F_p=1-e^{(as^2+bs+c)}$, where $s = \log_{10}^{\text{sweeps}}$, and $a$, $b$, and $c$ are constants, with $b$ being negative. Therefore, as sweeps become sufficiently large, $F_p$ tends to approach 1. Under 500 Gs, the constants for direct spin-state readout are $a = -0.31$, $b = 1.78$, and $c = -3.47$, while for the traditional readout, they are $a = -0.33$, $b = 1.45$, and $c = -1.28$.
(b) Evolution of fidelity $F_p$ as a function of experimental time under 500 Gs, which can be fitted using the function $F_p=1-e^{(a(\iota-\delta)^2+b(\iota-\delta)+c)}$, where $\iota=\log_{10}^{t}$ and $\delta$ is a constant determined by the duration of the single experiment. In the direct spin-state readout method, a 2500-ns laser pulse is used for direct spin-state readout, while in the general method, before laser pulse readout, a series of MW (2785 ns) and RF (156169 ns, 167389 ns) operations are required to manipulate the spin. The $\delta$ values for direct spin-state readout and traditional readout are 4.43 and 5.60, respectively.}
\end{figure} 

In practical experiments, the most significant factor contributing to the fidelity loss is the shot noise $\sigma_M$. Due to the photon counts follow a Poisson distribution, it is correlated with the photon signal, i.e $\sigma_M=\left[\sqrt{m^1}, \cdots\, \sqrt{m^i}, \cdots\, \sqrt{m^n}\right]^T$. This loss can be mitigated by increasing the number of measurement sweeps. Here, we conducted simulations based on experimental data to analyze the relationship between fidelity, the number of sweeps, and the total experiment time. The specific simulation process is as follows.

 Initially, we perform measurements to obtain the photon time traces for the four basis states through $10^9$ measurements. These traces are denoted as $\left\{L_{0\uparrow}, L_{0\downarrow}, L_{1\uparrow}, L_{1\downarrow}\right\}$, and we assume them to be error-free. The total number of measurements for this step is marked as $S1$.  In the second step, we randomly generate a set of target samples represented as $(c_{0\uparrow}^i, c_{0\downarrow}^i, c_{1\uparrow}^i, c_{1\downarrow}^i)$. Next, we generate theoretical data, $m^i_{\rm th}$, using the following equation: $m^i_{\rm th}=\frac{S1}{S2}c_{0\uparrow}^i l_{0\uparrow}^i+c_{0\downarrow}^i l_{0\downarrow}^i+c_{1\uparrow}^i l_{1\uparrow}^i+c_{1\downarrow}^i l_{1\downarrow}^i$. Here, $S2$  represents the number of experiment sweeps. To simulate the real experimental scenario, we add photon counting statistics noise, $\delta m^i_{\rm th}$, to the theoretical data, $m^i_{\rm th}$, resulting in experimental data, $m^i_{\rm exp}$. The noise, $\delta m^i_{\rm th}$, is drawn from a Gaussian distribution and falls within the range $[-\sqrt{m^i_{\rm th}}, \sqrt{m^i_{\rm th}}]$. By changing the value of $S1$, we can generate experimental data at different numbers of sweeps, which allows us to study the relationship between fidelity, sweeps, and experiment time. Finally, we use the generated photon time trace and experimental data, which have been normalized based on their maximum values, as inputs to an optimization model for analysis.

The evolution of fidelity $F_p$ as a function of the number of sweeps under a magnetic field strength of 500 Gs is depicted in Fig. \ref{fig:fig3}(a). When the fidelity level is not very high, the direct readout method requires fewer sweeps, especially when fidelity is less than 0.90. As the number of sweeps increases, the operation time of the general method grows linearly. In contrast, the direct readout method has a shorter readout time because it only includes the laser component and no additional operational time. Consequently, for the same number of sweeps, the direct readout method requires less time.

To further elucidate the relationship between fidelity and readout time, Fig. \ref{fig:fig3}(b) presents the corresponding data. When compared to the conventional general method, our approach demonstrates a substantial enhancement in time efficiency. For instance, when striving for a fidelity of 0.95, the direct readout method requires only $6.83\times 10^8$ ns, whereas the general method demands a significantly longer time of $2.24\times10^{10}$ ns. Therefore, in this particular scenario, our method accelerates the experimental process by a factor of 32. This time-saving achievement is primarily realized by reducing the number of essential spin operations, making it particularly advantageous for systems characterized by shorter decoherence times.

\section{dependence on magnetic field}  
\begin{figure}[H] 
	\includegraphics[width=0.5\textwidth]{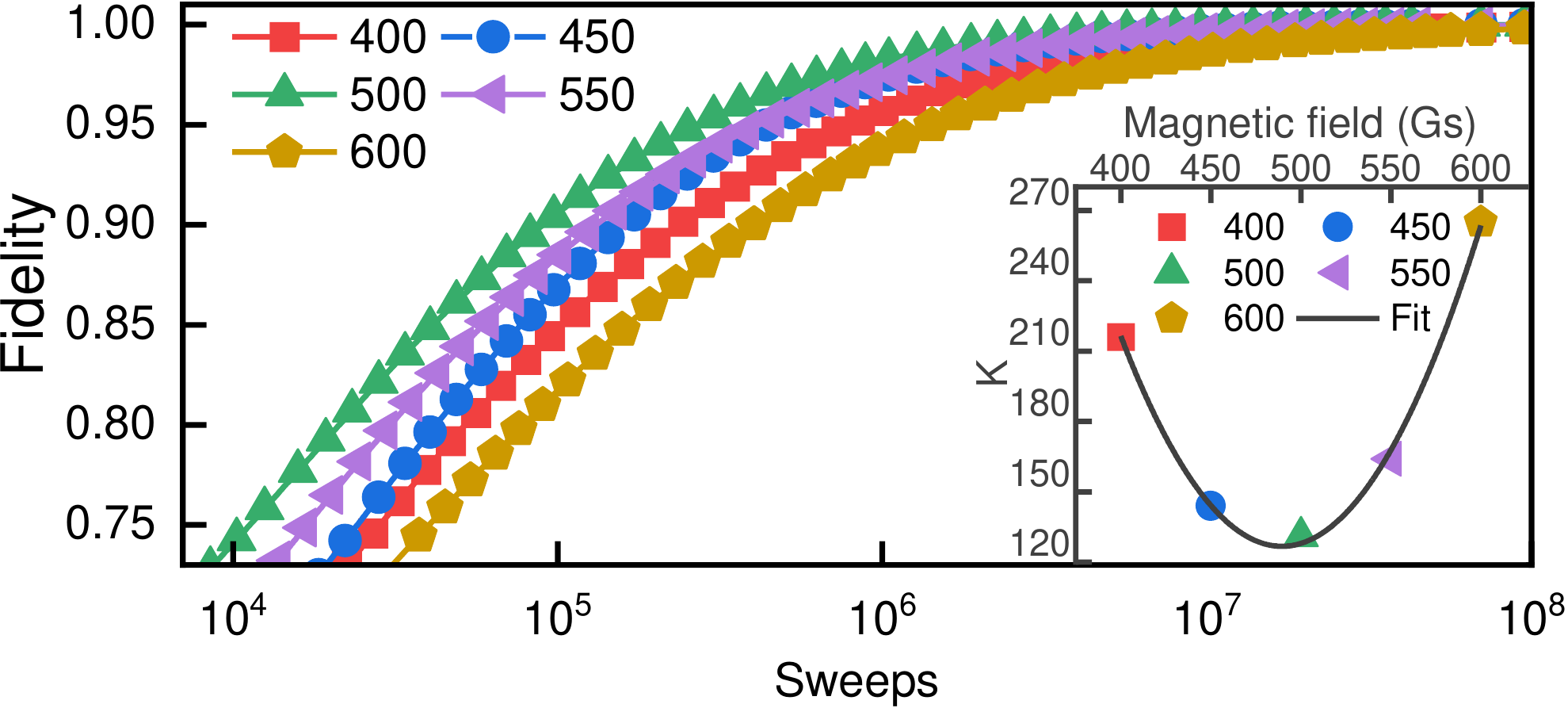}% Here is how to import EPS art
	\caption{\label{fig:fig4}Evolution of fidelity $F_p$ as a function of the sweeps and readout time for different fields. The simulate result are based on the photon time traces of $10^8$ measurements.}
\end{figure}

The ESLAC phenomenon is a crucial aspect of our method. Optical excitation at ESLAC results in state-selective spin mixing between the NV electron and nuclear spin. To gain deeper insights into our approach, we conducted measurements of the photon time trace under five different magnetic fields, comprising $10^8$ measurements each. Fig. \ref{fig:fig4} presents simulation data for these various magnetic field strengths, illustrating the relationship between fidelity $F_p$ and sweeps. The evolution of  $F_p$ as a function of sweeps can be well-fitted using the functions $F_p=1-e^{(as^2+bs+c)}$ and $F_p=1-e^{(a(\iota-\delta)^2+b(\iota-\delta)+c)}$. In the field regime close to 500 Gs (ESLAC point), we observe an optimal experimental effect where achieving the same fidelity level requires the fewest sweeps and the shortest time. This phenomenon arises due to the excited state electron-nuclear spin flip-flop process, induced by hyperfine interactions.

The spin flip frequency is closely related to the magnetic field strength, and it reaches an extreme value at approximately 500 Gs. Under the ESLAC conditions, the non-eigenstate $\left|0,\downarrow \right\rangle$ completely transforms into $\left|1,\uparrow \right\rangle$ and then back again, resulting in the most substantial differences in the photon time traces among the four initial states. In fact, even minor state mixing in the excited state can lead to noticeable distinctions in the photon time traces.

In the evaluation of photon time traces within the spin readout mechanism, we examine the worst-case noise magnification ratio $\kappa$\cite{PhysRevLett.105.010404,PhysRevA.94.052327}, which exhibits a quadratic correlation with the magnetic field, as shown in Fig. \ref{fig:fig4}. This observed change pattern aligns with the experimental results.

\section{discussion and conclusion}
In summary, we have effectively showcased a novel approach for achieving single-shot readout of nuclear spins linked to room-temperature NV centers in diamond. Our method centers around the direct analysis of photon time traces, with a specific emphasis on the discernible characteristics present during the ESLAC. When contrasted with conventional techniques, our approach introduces a substantial time-saving advantage, reducing the necessary time by a factor of 32, thanks to the elimination of spin operations prior to laser readout. While the ESLAC point represents the optimal condition for our method, it can also be applied to read out spin states whenever discernible differences in photon time traces are present.

 This method bestows benefits akin to those derived from diagonal element tomography experiments, and it can also be effectively applied to full-state tomography experiments. In state tomography experiments, it mitigates the requirement for intricate nuclear spin manipulation, offering a clear advantage in scenarios where preserving high fidelity poses challenges due to short decoherence times. This approach streamlines the experimental setup while retaining its capacity to uphold fidelity, rendering it a valuable tool for applications in quantum information processing and other domains that involve nuclear spin readout and manipulation.
 
\section{acknowledgments}
This work was supported by the Fundamental Research Funds for the Central Universities (Grant No. 226-2023-00137), the National Natural Science Foundation of China (Grant Nos. 92265114, 92265204, 12104213), the National Key Research and Development Program of China (2019YFA0308100) and the Innovation Program for Quantum Science and Technology (Grant No. 2021ZD0302200).

\appendix    
\section{simulation of time trace}

\begin{figure}[ht!]
    \centering
    \includegraphics[width=0.5\textwidth]{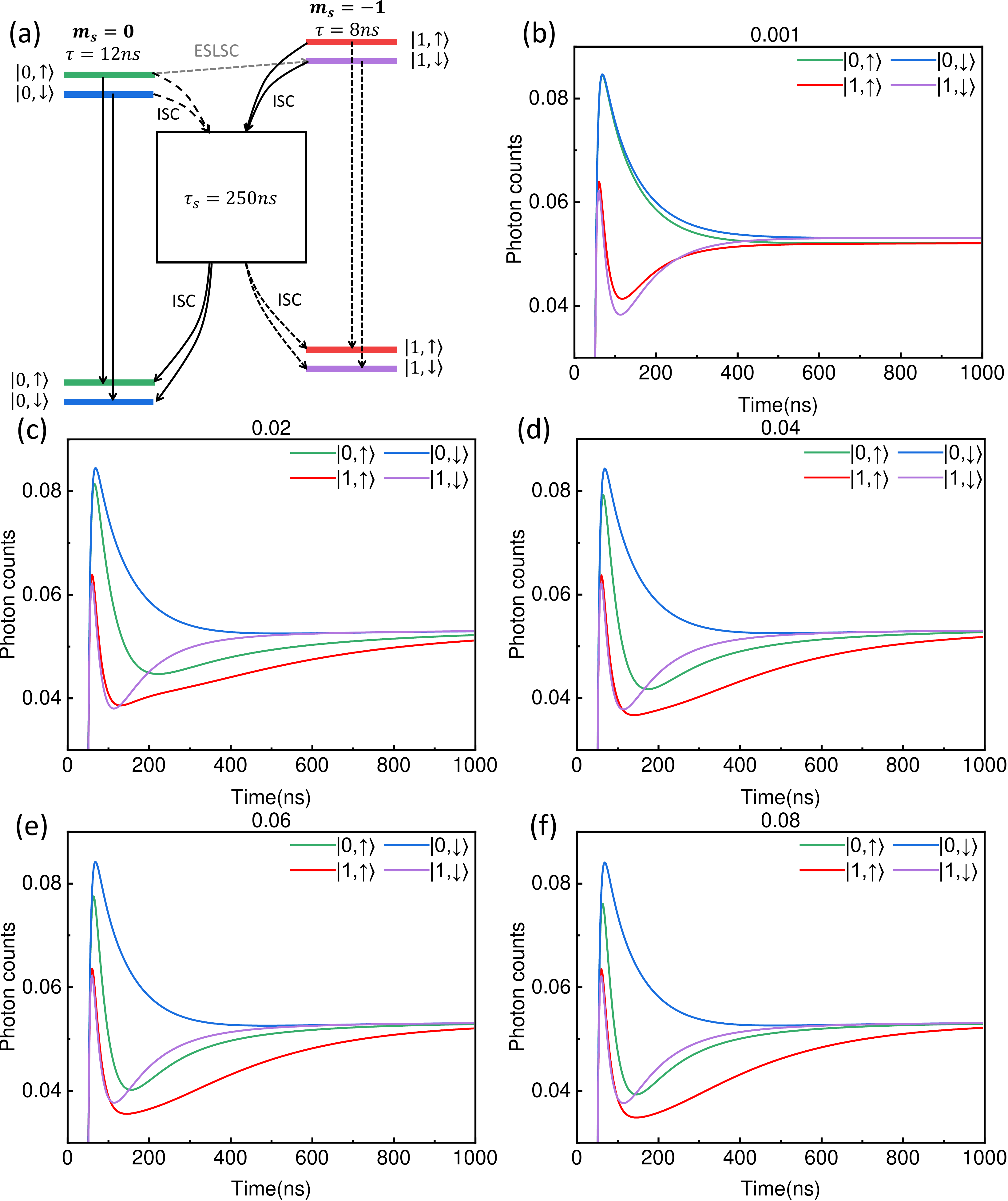}
    \caption{\label{fig:fig5} (a) Polarization mechanism of electron and nuclear spin. (b)-(f) Simulation results with different ESLAC rate 0.001, 0.02, 0.04, 0.06 and 0.08, respectively.  }
\end{figure}

\begin{figure}[ht!] 
    \includegraphics[width=0.5\textwidth]{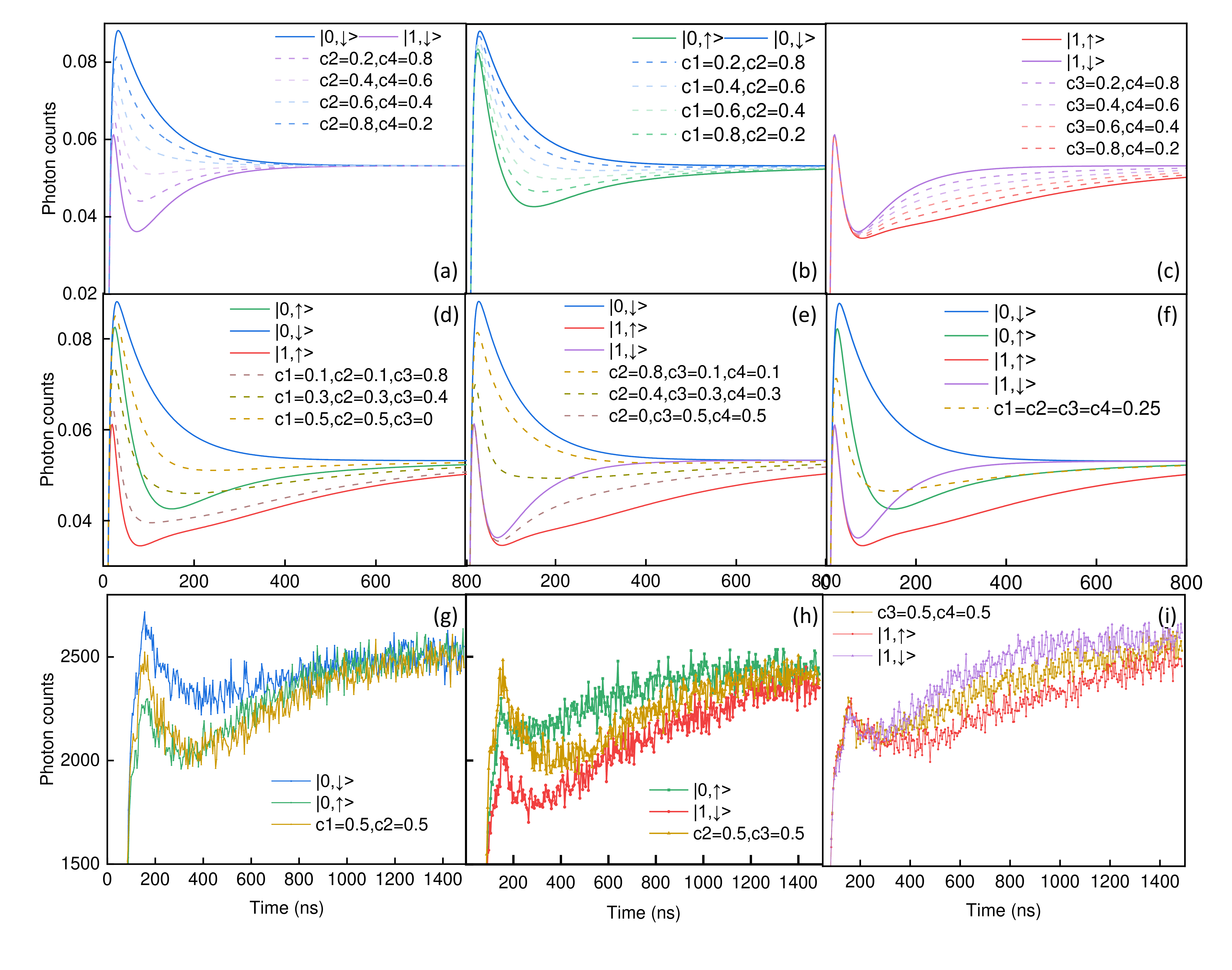}
    \caption{\label{fig:fig6}(a)-(c) illustrate the superposition states of two basis states, (d)-(e) show the superposition states of three basis states, (f) displays an arbitrary superposition state. The basis state is represented by solid lines, and the superposition states are depicted with dashed lines. (g)-(i) experiment result with $10^7$ measurements.}
\end{figure}

At a magnetic field strength of 500 Gs, a noteworthy phenomenon called ESLAC emerges (see Fig. \ref{fig:fig5}(a)), which will influence the fluorescence intensities among the four energy levels, namely $\left|0,\uparrow\right\rangle$, $\left|0,\downarrow\right\rangle$, $\left|1,\uparrow\right\rangle$, and $\left|1,\downarrow\right\rangle$. Here, we provided numerical simulations of the time trace with different ESLAC rates (see Fig. \ref{fig:fig5}). Furthermore, from the simulation of arbitrary superposition states in Fig. \ref{fig:fig6}, it can be observed that an arbitrary state is straightforwardly the superposition of the basic states.

% \begin{figurehere}
% \centering
% \includegraphics[width=50mm]{fig5}
% \end{figurehere}

\section{two-qubit full state tomography}\label{app2}
The density matrix $\rho$ of two-qubit entanglement is as followes:
\begin{small}
   \begin{equation}\label{b1}
    {\rho}_0=\left[\begin{array}{ccc}
        c_{0\uparrow}&\dotsc&a_{\left|0\uparrow\right\rangle\left\langle 1\downarrow\right|}+j b_{\left|0\uparrow\right\rangle\left\langle 1\downarrow\right|}\\
        a_{\left|0\uparrow\right\rangle\left\langle 0\downarrow\right|}-j b_{\left|0\uparrow\right\rangle\left\langle 0\downarrow\right|}&\dotsc&a_{\left|0\downarrow\right\rangle\left\langle 1\downarrow\right|}+j b_{\left|0\downarrow\right\rangle\left\langle 1\downarrow\right|}\\
        a_{\left|0\uparrow\right\rangle\left\langle 1\uparrow\right|}-j b_{\left|0\uparrow\right\rangle\left\langle 1\uparrow\right|}&\dotsc&a_{\left|1\uparrow\right\rangle\left\langle 1\downarrow\right|}+j b_{\left|1\uparrow\right\rangle\left\langle 1\downarrow\right|}\\
        a_{\left|0\uparrow\right\rangle\left\langle 1\downarrow\right|}-j b_{\left|0\uparrow\right\rangle\left\langle 1\downarrow\right|}&\dotsc&c_{1\downarrow}\\
        \end{array}
        \right].
\end{equation} 
\end{small}

The setup for measure diagonal elements $c_{0\uparrow},\ c_{0\downarrow},\ c_{1\uparrow}$ and $c_{1\downarrow}$ are explanated detaily in the second part of the main text. The  $a_{\left|0\uparrow\right\rangle\left\langle 0\downarrow\right|,\ \left|0\uparrow\right\rangle\left\langle 1\uparrow\right|,\ \left|0\uparrow\right\rangle\left\langle 1\downarrow\right|,\ \left|0\downarrow\right\rangle\left\langle 1\uparrow\right|,\ \left|0\downarrow\right\rangle\left\langle 1\downarrow\right|,\ \left|1\uparrow\right\rangle\left\langle 1\downarrow\right|}$ and $b_{\left|0\uparrow\right\rangle\left\langle 0\downarrow\right|,\ \left|0\uparrow\right\rangle\left\langle 1\uparrow\right|,\ \left|0\uparrow\right\rangle\left\langle 1\downarrow\right|,\ \left|0\downarrow\right\rangle\left\langle 1\uparrow\right|,\ \left|0\downarrow\right\rangle\left\langle 1\downarrow\right|,\ \left|1\uparrow\right\rangle\left\langle 1\downarrow\right|}$ are denote the real and imaginary part of off-diagonal elements $\left|0\uparrow\right\rangle\left\langle 0\downarrow\right|,\ \left|0\uparrow\right\rangle\left\langle 1\uparrow\right|,\ \left|0\uparrow\right\rangle\left\langle 1\downarrow\right|,\ \left|0\downarrow\right\rangle\left\langle 1\uparrow\right|,\ \left|0\downarrow\right\rangle\left\langle 1\downarrow\right|$ and $\left|1\uparrow\right\rangle\left\langle 1\downarrow\right|$ respectively.    
	
\begin{table}[ht]
    \resizebox{1\columnwidth}{!}{
    \begin{tabular}{ccc|ccc} 
        \hline
        off-diagonal & Readout Pulse & Counts&off-diagonal & Readout Pulse & Counts\\
        \hline
        \hline
        \multirow{4}{*}{$\left|0\uparrow\right\rangle\left\langle 0\downarrow\right|$} & $({\pi}_{RF_1}/2)_{\rm X}$ & $\rm X_1$&\multirow{4}{*}{$\left|0\downarrow\right\rangle\left\langle 1\uparrow\right|$}& $\   {\pi}_{FR_2}-({\pi}_{MW_2}/2)_{\rm X}$ & $\rm X_1\   $\\
        & $({\pi}_{RF_1})_{\rm -X}$ & $\rm X_2$&&$\   {\pi}_{RF_2}-({\pi}_{MW_2}/2)_{\rm -X}\   $ & $\rm X_2$\\
        & $({\pi}_{RF_1})_{\rm Y}$ & $\rm Y_1$&&$\   {\pi}_{RF_2}-({\pi}_{MW_2}/2)_{\rm Y}\   $ & $\rm Y_1$\\
        & $({\pi}_{RF_1})_{\rm -Y}$ & $\rm Y_2$&&$\   {\pi}_{RF_2}-({\pi}_{MW_2}/2)_{\rm -Y}\   $ & $\rm Y_2$\\
        \hline
        \multirow{4}{*}{$\left|0\uparrow\right\rangle\left\langle 1\uparrow\right|$} & $\   {\pi}_{RF_2}-{\pi}_{MW_2}-({\pi}_{RF_1}/2)_{\rm X}$ & $\rm X_1\   $&\multirow{4}{*}{$\left|0\downarrow\right\rangle\left\langle 1\downarrow\right|$}& $\   ({\pi}_{MW_2}/2)_{\rm X}$ & $\rm X_1\   $\\
        & $\   {\pi}_{RF_2}-{\pi}_{MW_2}-({\pi}_{RF_1}/2)_{\rm -X}\   $ & $\rm X_2$&& $\   ({\pi}_{MW_2}/2)_{\rm -X}\   $ & $\rm X_2$\\
        & $\   {\pi}_{RF_2}-{\pi}_{MW_2}-({\pi}_{RF_1}/2)_{\rm Y}\   $ & $\rm Y_1$&	& $\   ({\pi}_{MW_2}/2)_{\rm Y}\   $ & $\rm Y_1$\\
        & $\   {\pi}_{RF_2}-{\pi}_{MW_2}-({\pi}_{RF_1}/2)_{\rm -Y}\   $ & $\rm Y_2$&& $\   ({\pi}_{MW_2}/2)_{\rm -Y}\   $ & $\rm Y_2$\\
        \hline
        \multirow{4}{*}{$\left|0\uparrow\right\rangle\left\langle 1\downarrow\right|$} & $\   {\pi}_{MW_2}-({\pi}_{RF_1}/2)_{\rm X}$ & $\rm X_1\   $&\multirow{4}{*}{$\left|1\uparrow\right\rangle\left\langle 1\downarrow\right|$}& $\   ({\pi}_{RF_1}/2)_{\rm X}-{\pi}_{MW_2}$ & $\   \rm X_1\   $\\
        & $\   {\pi}_{MW_2}-({\pi}_{RF_1}/2)_{\rm -X}\   $ & $\rm X_2$&& $\   ({\pi}_{RF_1}/2)_{\rm -X}-{\pi}_{MW_2}\   $ & $\   \rm X_2$\\
        & $\   {\pi}_{MW_2}-({\pi}_{RF_1}/2)_{\rm Y}\   $ & $\rm Y_1$&& $\   ({\pi}_{RF_1}/2)_{\rm Y}-{\pi}_{MW_2}\   $ & $\   \rm Y_1$\\
        & $\   {\pi}_{MW_2}-({\pi}_{RF_1}/2)_{\rm -Y}\   $ & $\rm Y_2$&& $\   ({\pi}_{RF_1}/2)_{\rm -Y}-{\pi}_{MW_2}\   $ & $\   \rm Y_2$\\
        \hline
    \end{tabular}}
    \caption{\label{table3} 
        In this sequence, the ${\pi}_{MW_1}$, ${\pi}_{MW_2}$, ${\pi}_{RF_1}$, and ${\pi}_{RF_2}$ represent $\pi$-rotations applied to the respective transitions, as illustrated in Fig. 2 in main text, and the subscript $\rm X, -X, Y, -Y$ represent the four different phase of MW and RF pluse. }
\end{table}
For measure off-diagonal element, we need apply microwave(MW) or radio frequencies(RF) sequence to transform off-diagonal element into diagonal, then readout by laser. The off-diagonal experimental sequence is shown in \ref{table3}.

Next, take off-diagonal element $\left|0\uparrow\right\rangle\left\langle 1\downarrow\right|$ as an example for a detailed introduction. Firstly, we apply a $\rm MW_2\ \pi$ pulse, and the partial density matrix evolves as follows:
\begin{small}
    \begin{equation}\label{b2}
    \begin{split}
        \rho&=\left[\begin{array}{cccc}
        1&0&0&0\\
        0&0&0&-j\\
        0&0&1&0 \\
        0&-j&0&0\\
    \end{array}\right] {\rho}_0 \left[\begin{array}{cccc}
        1&0&0&0\\
        0&0&0&j\\
        0&0&1&0 \\
        0&j&0&0\\
    \end{array}\right]\\&=\left[\begin{array}{cccc}
        c_{0\uparrow}&ja_{\left|0\uparrow\right\rangle\left\langle 1\downarrow\right|}- b_{\left|0\uparrow\right\rangle\left\langle 1\downarrow\right|}&\dots&\dots\\
        -ja_{\left|0\uparrow\right\rangle\left\langle 1\downarrow\right|}+b_{\left|0\uparrow\right\rangle\left\langle 1\downarrow\right|}&c_{1\downarrow}&\dots &\dots\\
        \dots&\dots&c_{1\uparrow}&\dots \\
        \dots&\dots&\dots&c_{0\downarrow}\\ \end{array}
    \right].
    \end{split}   
\end{equation}
\end{small}

Then applying four $\rm RF_1\ \pi/2$ pulse with four different phases, the evolution of 2 $\times$ 2 density matrices $\rho_{1,2,3,4}^{sub}$ go as follows:
\begin{small}
\begin{equation}\label{b3}
\begin{split}
     \rho_0^{sub}&=\left[\begin{array}{cc}
        c_{0\uparrow}&ja_{\left|0\uparrow\right\rangle\left\langle 1\downarrow\right|}- b_{\left|0\uparrow\right\rangle\left\langle 1\downarrow\right|}\\
        -ja_{\left|0\uparrow\right\rangle\left\langle 1\downarrow\right|}+b_{\left|0\uparrow\right\rangle\left\langle 1\downarrow\right|}&c_{1\downarrow}\\
        \end{array}
        \right],\\
    \rho_1^{sub}&=\left[\begin{array}{cc}
            1&-j\\
            -j&1\\
        \end{array}\right]
        \rho_0^{sub}\left[\begin{array}{cc}
            1&j\\
            j&1\\
        \end{array}
        \right]\\
        &=\left[\begin{array}{cc}
            \frac{c_{0 \uparrow}+c_{1 \downarrow}}{2}-a_{\left|0\uparrow\right\rangle\left\langle 1\downarrow\right|}&\frac{j}{2}(c_{0 \uparrow}-c_{1 \downarrow})-b_{\left|0\uparrow\right\rangle\left\langle 1\downarrow\right|}\\
            -\frac{j}{2}(c_{0 \uparrow}-c_{1 \downarrow})+b_{\left|0\uparrow\right\rangle\left\langle 1\downarrow\right|}&\frac{c_{0 \uparrow}+c_{1 \downarrow}}{2}+a_{\left|0\uparrow\right\rangle\left\langle 1\downarrow\right|}\\
        \end{array}
        \right],\\
    \rho_2^{sub}&=\left[\begin{array}{cc}
            1&j\\
            j&1\\
        \end{array}
        \right]\rho_0^{sub}\left[\begin{array}{cc}
            1&-j\\
            -j&1\\
        \end{array}
        \right]\\
        &=\left[\begin{array}{cc}
            \frac{c_{0 \uparrow}+c_{1 \downarrow}}{2}+a_{\left|0\uparrow\right\rangle\left\langle 1\downarrow\right|}&-\frac{j}{2}(c_{0 \uparrow}-c_{1 \downarrow})-b_{\left|0\uparrow\right\rangle\left\langle 1\downarrow\right|}\\
            \frac{j}{2}(c_{0 \uparrow}-c_{1 \downarrow})-b_{\left|0\uparrow\right\rangle\left\langle 1\downarrow\right|}&\frac{c_{0 \uparrow}+c_{1 \downarrow}}{2}-a_{\left|0\uparrow\right\rangle\left\langle 1\downarrow\right|}\\
        \end{array}
        \right],\\
    \rho_3^{sub}&=\left[\begin{array}{cc}
            1&-1\\
            1&1\\
        \end{array}
        \right]\rho_0^{sub}\left[\begin{array}{cc}
            1&1\\
            -1&1\\
        \end{array}
        \right]\\
        &=\left[\begin{array}{cc}
            \frac{c_{0 \uparrow}+c_{1 \downarrow}}{2}+b_{\left|0\uparrow\right\rangle\left\langle 1\downarrow\right|}&\frac{1}{2}(c_{0 \uparrow}-c_{1 \downarrow})+j a_{\left|0\uparrow\right\rangle\left\langle 1\downarrow\right|}\\
            \frac{(c_{0 \uparrow}-c_{1 \downarrow})}{2}-ja_{\left|0\uparrow\right\rangle\left\langle 1\downarrow\right|}&\frac{c_{0 \uparrow}+c_{1 \downarrow}}{2}-b_{\left|0\uparrow\right\rangle\left\langle 1\downarrow\right|}\\
        \end{array}
        \right],\\
    \rho_4^{sub}&=\left[\begin{array}{cc}
            1&1\\
            -1&1\\
        \end{array}
        \right]\rho_0^{sub}\left[\begin{array}{cc}
            1&-1\\
            1&1\\
        \end{array}
        \right]\\
        &=\left[\begin{array}{cc}
            \frac{c_{0 \uparrow}+c_{1 \downarrow}}{2}-b_{\left|0\uparrow\right\rangle\left\langle 1\downarrow\right|}&-\frac{(c_{0 \uparrow}-c_{1 \downarrow})}{2}+ja_{\left|0\uparrow\right\rangle\left\langle 1\downarrow\right|}\\
            -\frac{(c_{0 \uparrow}-c_{1 \downarrow})}{2}-ja_{\left|0\uparrow\right\rangle\left\langle 1\downarrow\right|}&\frac{c_{0 \uparrow}+c_{1 \downarrow}}{2}+b_{\left|0\uparrow\right\rangle\left\langle 1\downarrow\right|}\\
        \end{array}
        \right].
\end{split}
\end{equation}
\end{small}

And now, the off-diagonal elements have been transfered to the elements, which can be optical readout. Following a laser readout, we get:
\begin{small}
   \begin{equation}\label{b4}
\begin{split}
  X_1&=(\frac{c_{0\uparrow}+c_{1\downarrow}}{2}-a_{\left|0\uparrow\right\rangle\left\langle 1\downarrow\right|})L_{0\uparrow}\\
  &+(\frac{c_{0\uparrow}+c_{1\downarrow}}{2}+a_{\left|0\uparrow\right\rangle\left\langle 1\downarrow\right|})L_{0 \downarrow}+c_{1\uparrow}L_{1 \uparrow}+c_{0 \downarrow}L_{1 \downarrow},
\end{split}
\end{equation}

\begin{equation}\label{b5}
\begin{split}
  X_2&=(\frac{c_{0\uparrow}+c_{1\downarrow}}{2}+a_{\left|0\uparrow\right\rangle\left\langle 1\downarrow\right|})L_{0\uparrow}\\
  &+(\frac{c_{0\uparrow}+c_{1\downarrow}}{2}-a_{\left|0\uparrow\right\rangle\left\langle 1\downarrow\right|})L_{0 \downarrow}+c_{1\uparrow}L_{1 \uparrow}+c_{0 \downarrow}L_{1 \downarrow},\\
\end{split}
\end{equation}

\begin{equation}\label{b6}
\begin{split}
  Y_1&=(\frac{c_{0\uparrow}+c_{1\downarrow}}{2}+b_{\left|0\uparrow\right\rangle\left\langle 1\downarrow\right|})L_{0\uparrow}\\
  &+(\frac{c_{0\uparrow}+c_{1\downarrow}}{2}-b_{\left|0\uparrow\right\rangle\left\langle 1\downarrow\right|})L_{0 \downarrow}+c_{1\uparrow}L_{1 \uparrow}+c_{0 \downarrow}L_{1 \downarrow},
\end{split}
\end{equation}

\begin{equation}\label{b7}
\begin{split}
  Y_2&=(\frac{c_{0\uparrow}+c_{1\downarrow}}{2}-b_{\left|0\uparrow\right\rangle\left\langle 1\downarrow\right|})L_{0\uparrow}\\
  &+(\frac{c_{0\uparrow}+c_{1\downarrow}}{2}+b_{\left|0\uparrow\right\rangle\left\langle 1\downarrow\right|})L_{0 \downarrow}+c_{1\uparrow}L_{1 \uparrow}+c_{0 \downarrow}L_{1 \downarrow},
\end{split}
\end{equation}

\begin{equation}\label{b8}
    a_{\left|0\uparrow\right\rangle\left\langle 1\downarrow\right|}=\frac{-X_1+X_2}{2(L_{0 \uparrow}-L_{0 \downarrow})},
\end{equation}
\begin{equation}\label{b9}
        b_{\left|0\uparrow\right\rangle\left\langle 1\downarrow\right|}=\frac{Y_1-Y_2}{2(L_{0 \uparrow}-L_{0 \downarrow})}.
\end{equation} 
\end{small}

\begin{figure}[H] 
	\includegraphics[width=\columnwidth]{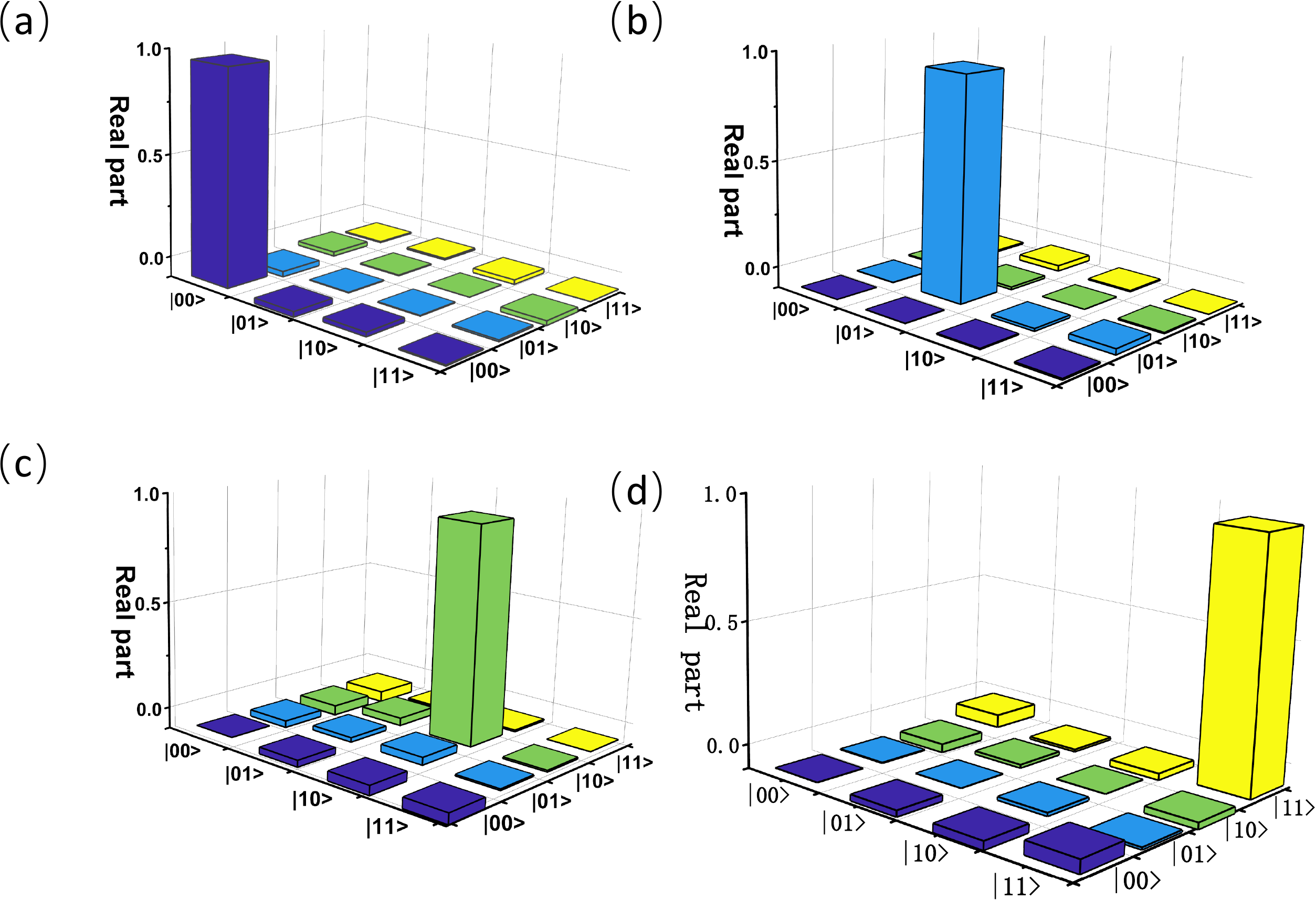}% Here is how to import EPS art
	\caption{\label{fig:fig7}(a)-(d) Tomography result of state $\left|0,\uparrow\right\rangle, \left|0,\downarrow\right\rangle, \left|1,\uparrow\right\rangle$, and $\left|1,\downarrow\right\rangle$ respectively. The fidelity of these four states is all above 0.99.}
\end{figure}

In this work, the all off-diagonal elements are calculated by the similar method, and the results of tomography as shown in Fig. \ref{fig:fig7}.
	
\bibliographystyle{unsrt}
%\bibliography{reference}
%merlin.mbs apsrev4-1.bst 2010-07-25 4.21a (PWD, AO, DPC) hacked
%Control: key (0)
%Control: author (8) initials jnrlst
%Control: editor formatted (1) identically to author
%Control: production of article title (-1) disabled
%Control: page (0) single
%Control: year (1) truncated
%Control: production of eprint (0) enabled
%

% \end{appendix}

\end{document}